\documentclass{aa}
\usepackage{graphicx,natbib}
\usepackage[varg]{txfonts}
\usepackage{color}

\newcommand{\av}[1]{\langle{#1}\rangle}

\newcommand{\Hm}{\langle B\rangle}
\newcommand{\Hz}{\langle B_z\rangle}

\newcommand{\Hq}{\av{B_{\rm q}}}
\newcommand{\Hsq}{\av{B^2}}
\newcommand{\Hzsq}{\av{B_z^2}}

\newcommand{\Hav}{\Hm_{\rm av}}

\newcommand{\Hzrms}{\Hz_{\rm rms}}

\newcommand{\Prot}{P_{\rm rot}}

\newcommand{\Zeeman}{\Delta\lambda_{\rm Z}}

\newcommand{\OC}{\ensuremath{{\rm O}-{\rm C}}}
\newcommand{\Feline}{Fe~{\sc ii}~$\lambda\,6149.2\,\AA$}
\newcommand{\Felineb}{Fe~{\sc ii/i}~$\lambda\,6147.7\,\AA$}
\newcommand{\Felinebb}{Fe~{\sc ii}~$\lambda\,6147.7\,\AA$}
\newcommand{\Ndline}{Nd~{\sc iii}~$\lambda\,6145\,\AA$}

\bibpunct[ ]{(}{)}{;}{a}{}{,}

\begin{document}

\title{The 10.5 year rotation period of the strongly magnetic\\
rapidly oscillating Ap star HD~166473\thanks{Based on observations
  collected at the Canada-France-Hawaii Telescope (CFHT), which is
  operated by the National Research Council of Canada, the Institut
  National des Sciences de l'Univers of the Centre National de la
  Recherche Scientifique of France, and the University of Hawaii; and
  on data products from
observations made with ESO Telescopes at the La Silla Paranal
Observatory under programmes 79.C-0170, 80.C-0032, 81.C-0034, and
  89.D-0383. The
operations at the Canada-France-Hawaii Telescope are conducted with
care and respect from the summit of Mauna Kea, which is a significant
cultural and historic site.}}

\author{G.~Mathys\inst{1}
  \and V.~Khalack\inst{2}
  \and J.~D.~Landstreet\inst{3,4}
}

\institute{European Southern Observatory,
  Alonso de Cordova 3107, Vitacura, Santiago, Chile\\\email{gmathys@eso.org}
\and
  D\'epartement de Physique et d'Astronomie, Universit\'e de Moncton,
  Moncton, NB, Canada E1A 3E9
\and
Department of Physics \& Astronomy, University of Western Ontario,
1151 Richmond Street, 
London, Ontario N6A 3K7, Canada 
\and
Armagh Observatory, College Hill, Armagh, BT61 9DG, Northern Ireland,
UK 
}

\date{Received $\ldots$ / Accepted $\ldots$}

\titlerunning{The 10.5 year rotation period of HD~166473}

\abstract{How magnetic fields contribute to the differentiation of the rotation
  rates of the Ap stars and affect the occurrence of non-radial
  pulsation in some of them are important open questions. Valuable
  insight can be gained into these questions by studying some of the
  most extreme examples of the processes at play. The super-slowly
rotating rapidly oscillating Ap (roAp) star HD~166473 is such an
example.}
{We performed the first accurate determination of its rotation
period, $\Prot=(3836\pm30)$\,d, from the analysis of 56 measurements
of the mean magnetic field modulus $\Hm$ 
  based on high-resolution spectra acquired between 1992 and 2019 at
  various observatories and with various instrumental
  configurations.}{We complemented this 
  analysis with the consideration of an inhomogeneous set of 21
  determinations of the mean longitudinal magnetic
  field $\Hz$ spanning the same time interval.}{This makes HD~166473
  one of only four Ap stars with a period longer than 10 years for
  which magnetic field measurements have been obtained over more than
  a full cycle. The
  variation curves of $\Hm$ and of $\Hz$ are well approximated by 
  cosine waves. The magnetic field of HD~166473 only seems to deviate
  slightly from axisymmetry, but it definitely involves a
  considerable non-dipolar component.}{Among the stars with rotation
  periods longer than 1000\,d for which magnetic field measurements
  with full phase coverage are available, HD~166473 has the 
  strongest field. Its magnetic field is also one of the strongest
  known among roAp stars. Overall, the magnetic properties of
  HD~166473 do not seem fundamentally distinct from those of the
  faster-rotating Ap stars. However, considering as a group the eight
  Ap stars that have accuractely determined periods longer than 1000\,d
  and whose magnetic variations have been characterised over a full
  cycle suggests that the angles between their magnetic and rotation
  axes tend to be systematically large.}
  

\maketitle

\section{Introduction}
\label{sec:intro}
The existence of a sizeable population of Ap stars that rotate
extremely slowly is now well established. According to
\citet{2017A&A...601A..14M}, stars with rotation periods ($\Prot$)
longer than one year may represent several percent of all Ap
stars. At present, at least 20 Ap stars that definitely have
$\Prot>1$\,yr are known. But statistical arguments indicate with a
high level of confidence that there exist many more Ap stars with such
long periods.

Consideration of the extremely slowly rotating Ap stars is important
to understand the origin and the evolution of the rotational
properties of Ap stars as a class, and more generally, of all upper
main sequence stars. Within that context, identifying dependencies or
correlations between the rotation rates and other properties of these
stars can potentially provide very valuable insight. As an example of
such a correlation, \citet{1997A&AS..123..353M} reported that very
strong magnetic fields ($\Hav\ga7.5$\,kG) are found only in stars with
rotation periods shorter than $\sim$150 days. This result was
subsequently confirmed, at a very high level of significance, by
\citet{2017A&A...601A..14M}. The notation $\Hav$ denotes the average
over a stellar 
rotation cycle of the mean magnetic field modulus ($\Hm$, the
line-intensity weighted average
over the visible stellar disk of the modulus of the magnetic
vector).

Until now, magnetic field measurements covering more than a full
rotation cycle with adequate sampling have been obtained for seven of
the 20 Ap stars known to have a rotation period longer than one
year. Such a complete phase coverage is essential to characterise the
main properties of the stellar magnetic field, including its geometric
structure. This characterisation is required to identify possible
correlations between the rotation rate, the magnetic field, and other
physical properties (e.g., mass, age, abundance anomalies and
inhomogeneities) of the Ap stars.

With regard to the connection between rotation and other physical
properties, the situation of the rapidly oscillating Ap
(roAp) stars is particularly intriguing. Of the 61 roAp stars
inventoried by \citet{2015MNRAS.452.3334S}, 14 show spectral lines
resolved into their magnetically split components \citep[see
Tables~1 and 2 of][]{2017A&A...601A..14M}. In other words,
the rate of occurrence of magnetically resolved lines in roAp stars
seems significantly higher than average -- no more than a few percent
of all Ap stars. While a few of the Ap stars with resolved magnetically split
lines may be short period stars whose rotation axis has
a low inclination to the line of sight, most of them are genuine slow
rotators \citep{2017A&A...601A..14M}. This suggests that slow rotators
may also be more frequent among roAp stars than among Ap stars in
general. As a matter of fact, of the 14 roAp stars with resolved
magnetically split lines, nine definitely have, or may plausibly have, a
rotation period longer than one year. Furthermore, at least two of the
roAp stars with unresolved lines are also suspected to have
$\Prot>1$\,yr: HD~101065 (Przybylski's star), for which
\citet{2018MNRAS.477.3791H} proposed a tentative, extrapolated value
$\Prot\approx188$\,yr, and HD~176232 (= 10~Aql), for which
\citet{1992A&A...256..169M} suggested that the period might be of the
order of years, and \citet{2017PASP..129j4203P} did not detect any
photometric variation over a time interval of 12 years. Based on these
considerations, it is not implausible 
that more than 15\% of the roAp stars could have rotation periods in
excess of one year -- a strikingly high fraction compared to the rate
of occurrence of extremely slow rotation in the entire population of
Ap stars. Whether this high fraction is indicative of a direct
connection between slow rotation and the occurrence of non-radial
pulsation, or if it
results from independent links between each of those two features and
other physical parameters of the stars where both occur simultaneously
(such as the mass) remains to be elucidated. 

Within this context, the A5p SrCrEu star \citep{2009A&A...498..961R}
HD~166473 (= V694~CrA) is a particularly 
interesting specimen in several respects. \citet{1987MNRAS.226..187K}
discovered that HD~166473 is a roAp star, showing low amplitude
photometric variations with periods between 8.8 and
9.1\,min. \citet{1997A&AS..123..353M} reported the observation of
resolved magnetically split lines in the spectrum of this star. The
first measurements of its mean longitudinal magnetic field ($\Hz$, the line
intensity weighted average over the stellar disk of the component of
the magnetic vector along the line of sight) were carried out by
\cite{1997A&AS..124..475M}. \citet{2003MNRAS.343L...5K} reported
observations of radial velocity variations with the pulsation
frequencies, which were analysed in greater detail by
\cite{2007MNRAS.380..181M}. These authors presented additional
measurements of the mean magnetic field modulus, and combined them
with determinations of other magnetic field moments \citep[which were
eventually published by][]{2017A&A...601A..14M} to infer that the
rotation period of HD~166473 must be of the order of 10\,yr. More
recently, \citet{2014IAUS..302..274S} reported a value of
$\Prot=3514$\,d for the rotation period. This value appears to have
  been estimated from the measurements published by
  \citet{2007MNRAS.380..181M}, which span a time base of 3631\,d,
  and from six additional determinations of $\Hm$, derived from
    spectra recorded at unspecified epochs spread over a time interval
    of similar length.

From the studies listed above, it definitely emerges that the rotation
period of HD~166473 cannot be significantly shorter than 10\,yr and
that this star has a strong magnetic field, with a value of the order
of 7 kG for the average of the mean magnetic field modulus over a
rotation cycle. This field strength locates it very close to the upper
limit that was identified for the fields of Ap stars with rotation
periods longer than $\sim$150\,d. Moreover, among the roAp stars,
until now, stronger fields have only been observed in
two: HD~154708 \citep[24.5\,kG,][]{2006MNRAS.372..286K} and HD~92499
\citep[8.2\,kG,][]{2010MNRAS.404L.104E}. This is significant with
respect to the relationship between the magnetic field and pulsation
in roAp stars. For a description of the multiple facets of this
relationship, see e.g. \citet{2006MNRAS.372..286K}.

Thus, in terms of its field
strength, HD~166473 is one of the most extreme examples both of the
population of super-slowly rotating Ap stars and of the class of the
roAp stars. In many physical contexts, such extreme examples are known
to provide very valuable insight into the processes at
play. This represents a strong motivation to study HD~166473 in
detail.

\begin{table}[ht!]
\caption{Mean magnetic field modulus measurements.}
\label{tab:Hm}
\centering
\begin{tabular}{lrc}
\hline\hline\\[-4pt]
  \multicolumn{1}{c}{JD}&\multicolumn{1}{c}{$\Hm$}&Configuration\\
  &\multicolumn{1}{c}{(G)}&\\[4pt]
\hline\\[-4pt]
 2448788.758&8551&ESO CAT + CES LC\\
 2448790.817&8444&ESO CAT + CES LC\\
 2448841.672&8454&ESO CAT + CES LC\\
 2448911.512&8540&ESO CAT + CES LC\\
 2449079.772&8412&ESO CAT + CES LC\\
 2449100.913&8305&ESO CAT + CES LC\\
 2449130.878&8138&ESO CAT + CES LC\\
 2449160.832&8148&ESO CAT + CES LC\\
 2449212.719&7991&ESO CAT + CES LC\\
 2449215.670&7964&ESO CAT + CES LC\\
 2449298.510&7878&ESO CAT + CES SC\\
 2449301.510&7970&ESO CAT + CES SC\\
 2449419.877&7645&ESO CAT + CES LC\\
 2449437.843&7735&ESO CAT + CES SC\\
 2449456.858&7552&ESO CAT + CES LC\\
 2449494.717&7297&ESO CAT + CES LC\\
 2449531.786&7341&ESO CAT + CES LC\\
 2449793.849&6686&ESO CAT + CES LC\\
 2449829.883&6651&ESO CAT + CES LC\\
 2449853.877&6699&ESO CAT + CES LC\\
 2449881.848&6680&ESO CAT + CES LC\\
 2449908.774&6456&ESO CAT + CES LC\\
 2449947.536&6395&ESO CAT + CES LC (lower res.)\\
 2450132.885&5969&ESO CAT + CES LC\\
 2450149.872&5883&ESO CAT + CES LC\\
 2450171.896&5789&ESO CAT + CES LC\\
 2450231.800&5820&ESO CAT + CES LC\\
 2450522.835&5661&ESO CAT + CES LC\\
 2450702.538&5770&ESO CAT + CES LC\\
 2450900.853&6068&ESO CAT + CES LC\\
 2450971.819&6095&ESO CAT + CES LC\\
 2451042.606&6038&ESO CAT + CES LC\\
 2451084.582&6118&ESO CAT + CES LC\\
 2451326.017&6617&CFHT + Gecko\\
 2451326.024&6623&CFHT + Gecko\\
 2451600.858&7051&ESO 3.6\,m + CES VLC\\
 2451739.851&7565&CFHT + Gecko\\
 2451740.863&7487&CFHT + Gecko\\
 2452090.722&8450&ESO VLT UT2 + UVES\\
 2452189.506&8349&ESO VLT UT2 + UVES\\
 2452419.958&8790&CFHT + Gecko\\
 2453214.889&7557&CFHT + Gecko\\
 2454336.612&5748& ESO 3.6\,m + HARPS\\
 2454338.630&5709& ESO 3.6\,m + HARPS\\
 2454544.782&5761& ESO 3.6\,m + HARPS\\
 2454545.866&5754& ESO 3.6\,m + HARPS\\
 2454633.768&5745& ESO 3.6\,m + HARPS\\
 2454634.852&5766& ESO 3.6\,m + HARPS\\
 2454716.624&5794& ESO 3.6\,m + HARPS\\
 2456148.655&8518& ESO 3.6\,m + HARPS\\
 2456531.773&8540&CFHT + ESPaDOnS\\
 2456547.732&8487&CFHT + ESPaDOnS\\
 2456813.014&8203&CFHT + ESPaDOnS\\
 2457239.836&7278&CFHT + ESPaDOnS\\
 2457287.712&7170&CFHT + ESPaDOnS\\
 2458642.993&5803&CFHT + ESPaDOnS\\[4pt]
\hline
\end{tabular}
\end{table}

Here we present new determinations of the mean magnetic field modulus
and of the mean longitudinal magnetic field of HD~166473, based on
both dedicated observations and archive spectra. We derived for the
first time an accurate value of the rotation period of the star from the
complete set of existing $\Hm$ measurements. The observational data
and their analysis are presented in Sect.~\ref{sec:obs}, and the
determination of the stellar rotation period is described in
Sect.~\ref{sec:per}. In Sect.~\ref{sec:magfield}, we discuss the
constraints that the available magnetic data set on the field
geometry, and related surface inhomogeneities. Finally, we consider
the implications of the results of this analysis of HD~166473 within
the contexts of Ap star rotation and pulsation. 

\section{Observations and data analysis}
\label{sec:obs}
\subsection{Mean magnetic field modulus}
\label{sec:Hm}
All the mean field modulus values used in this analysis were
determined from the 
measured wavelength separation of the two magnetically split
components of the \Feline\ diagnostic line. The following formula was
applied to derive $\Hm$:
\begin{equation}
\lambda_{\rm r}-\lambda_{\rm b}=g\,\Zeeman\,\Hm\,.
\label{eq:Hm}
\end{equation}
In this equation, $\lambda_{\rm r}$ and $\lambda_{\rm b}$ are,
respectively, the wavelengths of the red and blue split line
components; $g$ is the Land\'e factor of the split level of the
transition ($g=2.70$; \citealt{1985aeli.book.....S}); 
$\Zeeman=k\,\lambda_0^2$, with
$k=4.67\,10^{-13}$\,\AA$^{-1}$\,G$^{-1}$; $\lambda_0=6149.258$\,\AA\
is the nominal wavelength of the considered transition.

As in many other Ap stars, the \Feline\ line in HD~166473 is blended
on the blue side with an unidentified rare earth line. However, this
blend is less severe than in other stars, so that its impact on the
precision and uniformity of the $\Hm$ measurements does not represent
a major concern (see below). In particular, it does not
introduce any significant 
ambiguity in the combination of past measurements with new ones,
unlike in the case of HD~965 \citep{2019A&A...629A..39M}. Thus,
the present analysis makes use of the following published measurements
of the mean magnetic field modulus of HD~166473:
23 measurements from \citet{1997A&AS..123..353M},
ten measurements from \citet{2017A&A...601A..14M},
and seven additional measurements from \citet{2007MNRAS.380..181M}.
These measurements were obtained with six different instrumental
configurations. Five of these configurations were described by
\citet{2017A&A...601A..14M}; for the sake of simplicity, we use the
same symbols as this author to identify them in
Fig.~\ref{fig:bmcurve}. One measurement of \citet{2007MNRAS.380..181M}
was based on a spectrum 
recorded with the Very Long Camera (VLC) of the ESO 
    Coud\'e Echelle Spectrograph (CES) fed by the ESO 3.6-m telescope
    \citep{2005CESVLC}. This configuration is quite different from
    those used by \citet{1997A&AS..123..353M} and
    \citet{2017A&A...601A..14M}. This difference was overlooked by
    \citet{2017A&A...601A..14M}.

Here, we present new $\Hm$ data at additional epochs, from the
analysis of the following high-resolution spectra recorded in natural
light:
\begin{itemize}
\item one spectrum recorded with the Gecko spectrograph fed by the
  Canada-France-Hawaii Telescope (CFHT). This spectrum was reduced in
  the same way as previous spectra obtained with the same
  configuration \citep[see Section 3 of][for
    details]{1997A&AS..123..353M};
  \item six spectra recorded with the ESPaDOnS spectrograph fed by the
    Canada-France-Hawaii Telescope (CFHT). These spectra were reduced
    using the dedicated software package Libre-ESpRIT
    \citep{1997MNRAS.291..658D}, which yields both the Stokes $I$
    spectrum and the Stokes $V$ circular polarisation spectrum. The
    Stokes $I$ spectra were normalised using the same procedure as in
    \citet{2017MNRAS.471..926K}; 
    \item one spectrum recorded with the Ultraviolet and Visible
        Echelle Spectrograph (UVES) fed by Unit Telescope 2 (UT2) of
        the ESO Very Large Telescope (VLT), retrieved from the ESO
        Archive;
        \item and eight spectra recorded with the High Accuracy Radial velocity
      Planet Searcher (HARPS) fed by the ESO 3.6-m telescope,
      retrieved from the  ESO Archive. One of these spectra was
      obtained with the polarimetric mode of HARPS
      \citep[HARPSpol;][]{2011Msngr.143....7P} and is further 
      discussed in Sect.~\ref{sec:Hz}. 
    \end{itemize}
For the HARPS and UVES observations, we used science grade pipeline
processed data available from the ESO
Archive.
% \footnote{http://archive.eso.org/wdb/wdb/adp/phase3\_spectral/form}.
The only additional 
processing that we carried out was a continuum normalisation of the
region ($\sim$100\,\AA\ wide) surrounding the \Feline\ diagnostic line
(except for the HARPSpol spectrum -- see below).  

The procedure actually used to measure the wavelengths $\lambda_{\rm
  b}$ and $\lambda_{\rm r}$ of the \Feline\ line is described in
detail by \citet{1992A&A...256..169M} and by
\citet{1997A&AS..123..353M}. We fitted three gaussians to the blend 
consisting of the split Fe~{\sc ii} doublet and the unidentified rare
earth line on its blue side. As stressed by
\citet{1997A&AS..123..353M}, this represents a very effective way to
disentangle the contribution of the rare earth blend from that of the two
Fe~{\sc ii} line components and to achieve consistent determinations
of the wavelengths of the latter. The usage of this approach also
ensures that, for a given star, the achievable precision in the
measurements of $\lambda_{\rm b}$ and $\lambda_{\rm r}$ 
is almost independent of the spectral resolution and of the S/N of the
observations acquired at different epochs, as long as both are
sufficiently high. The resulting consistency in the derived values of
$\Hm$ between  observations obtained at different epochs with
different instrumental configurations and the uniformity of the
uncertainties of these measurements are abundantly illustrated by the
phase diagrams of the mean magnetic field modulus variations for
numerous stars shown in Appendix~A of \citet{2017A&A...601A..14M}. For
each star, the scatter of the measurements about a smooth variation
curve is very uniform, independently of the measurement source. In
other words, the adopted measurement procedure is very robust. In
general, it is straightforward to ensure that all the measurements are
performed in a repeatable manner, even if they were carried out years
apart for different subsets of data. In only a few exceptional cases,
some additional caution may be required, and achieving sufficient
precision may require all the spectra of a given star to be measured
in a single batch to ensure consistency. Such exceptions typically
arise from the presence of an unusually strong or highly variable
rare earth blend to the \Feline\ line, combined with a rather weak or
very strong magnetic field. Examples include HD~965
\citep{2019A&A...629A..39M} and HD~318107
\citep[=~CoD~$-$32\,13074][]{2000A&A...364..689M,2017A&A...601A..14M}. The
case of HD~166473 is much less challenging.

Accordingly, there is no reason to expect the uncertainty of the
  new determinations 
of $\Hm$ presented here to be significantly different from that of the
previous measurements of \citet{1997A&AS..123..353M},
\citet{2007MNRAS.380..181M}, and \citet{2017A&A...601A..14M}. We shall
see in 
Sect.~\ref{sec:magfield} that the scatter of the measurements about
the variation curve of the mean magnetic field modulus is fully
consistent with this adopted value of the uncertainty,
80\,G.

\begin{table}
\caption{Mean longitudinal magnetic field measurements.}
\label{tab:Hz}
%\centering
\begin{tabular}{@{}@{\extracolsep{\fill}}crrl @{\extracolsep{0pt}}@{}}
\hline\hline\\[-4pt]
\multicolumn{1}{c}{JD}&\multicolumn{1}{c}{$\Hz$ (G)}&$\sigma_z$ (G)&Reference\\[4pt]
\hline\\[-4pt]
 2448782.707&$-$2036&105&\protect{\citet{1997A&AS..124..475M}}\\
 2448845.631&$-$2291&314&\protect{\citet{1997A&AS..124..475M}}\\
 2448846.696&$-$2110&198&\protect{\citet{1997A&AS..124..475M}}\\
 2449830.876&  245&129&\protect{\citet{2017A&A...601A..14M}}\\
 2449916.826&  711&119&\protect{\citet{2017A&A...601A..14M}}\\
 2449972.637&  841&213&\protect{\citet{2017A&A...601A..14M}}\\
 2450183.873& 1458&185&\protect{\citet{2017A&A...601A..14M}}\\
 2450294.775& 1679&133&\protect{\citet{2017A&A...601A..14M}}\\
 2450497.889& 1746&155&\protect{\citet{2017A&A...601A..14M}}\\
 2450616.882& 1822&116&\protect{\citet{2017A&A...601A..14M}}\\
 2454209.827& 2326& 44&\protect{\citet{2015A&A...583A.115B}}\\
 2454247.693& 2298& 54&\protect{\citet{2015A&A...583A.115B}}\\
 2454250.896& 2306& 45&\protect{\citet{2015A&A...583A.115B}}\\
 2454308.785& 2411& 45&\protect{\citet{2015A&A...583A.115B}}\\
 2456148.655&$-$1218&61&This paper (HARPSpol)\\
 2456531.773&$-$1741&74&This paper (ESPaDOnS)\\
 2456547.732&$-$1747&69&This paper (ESPaDOnS)\\
 2456813.014&$-$1495&55&This paper (ESPaDOnS)\\
 2457239.836&$-$446&37&This paper (ESPaDOnS)\\
 2457287.712&$-$292&45&This paper (ESPaDOnS)\\
 2458642.993& 1592&99&This paper (ESPaDOnS)\\[4pt]
\hline
\end{tabular}
\end{table}

The 16 new values of the mean magnetic field modulus that we derived
from the analysis of the additional spectra listed above are 
presented in Table~\ref{tab:Hm}. For the convenience of the reader,
this table also includes the 40 previously published
measurements. The
columns give, in order, the Heliocentric (or Barycentric, for HARPS)
Julian Date of mid-exposure, the value
$\Hm$ of the mean magnetic field modulus, and the instrumental
configuration with which the analysed spectrum was obtained. This
information had never been specified on an individual basis for some
of the previously published measurements. For more detailed
descriptions of the configurations used in past studies, see
\citet{2017A&A...601A..14M} and references therein.

\subsection{Mean longitudinal magnetic field}
\label{sec:Hz}
The first three measurements of the mean longitudinal magnetic field of
HD~166473 were performed by \citet{1997A&AS..124..475M} through the
analysis of spectra recorded in both circular polarisations with the
ESO Cassegrain Echelle Spectrograph (CASPEC) fed by the ESO 3.6-m
telescope. The value of $\Hz$ was determined from the wavelength
shifts of a sample of spectral lines between  
the two circular polarisations in each of these spectra, by application of the formula:
\begin{equation}
\lambda_{\rm R}-\lambda_{\rm L}=2\,\bar g\,\Zeeman\,\Hz\,,
\label{eq:Hz}
\end{equation}
where $\lambda_{\rm R}$ (resp. $\lambda_{\rm L}$) is the wavelength of
the centre of gravity of the line in right (resp. left) circular
polarisation and $\bar g$ is the effective Land\'e factor of the
transition. $\Hz$ is determined through a 
least-squares fit of the measured values of $\lambda_{\rm
  R}-\lambda_{\rm L}$ by a function of the form given above. The standard error
$\sigma_z$ that is derived from that
least-squares analysis is used as an estimate of the uncertainty
affecting the obtained value of $\Hz$. The same technique was
subsequently applied by \citet{2017A&A...601A..14M} to obtain $\Hz$
measurements at seven additional epochs from spectra recorded with the
same instrument and telescope combination. 

The catalogue of \citet{2015A&A...583A.115B} lists four more
values of $\Hz$ in HD~166473, based on observations carried out with
FORS-1 in its spectropolarimetric mode, fed by one of the Unit
Telescopes of the ESO VLT.  The interpretation of the FORS-1
spectropolarimetric data in terms of the mean longitudinal magnetic
field, which is described in detail by \citet{2015A&A...583A.115B},
rests on assumptions and approximations that are different from those
underlying Eq.~(\ref{eq:Hz}). The $\Hz$ values derived in that way may
not be fully consistent with those obtained from the above-described
analysis of the CASPEC data. Such disagreements between FORS-1 $\Hz$
measurements and determinations of this field moment through the
analysis of high-resolution spectropolarimetric observations are not
unusual \citep{2014A&A...572A.113L}. 

To complement the $\Hz$ values from the literature, here we present
seven new determinations of this magnetic field moment, based on the
analysis of circularly polarised spectra, one of them 
recorded with the polarimetric mode of HARPS at the ESO 3.6\,m
telescope and the other six with the ESPaDOnS spectropolarimetre
\citep{2006ASPC..358..362D} at the CFHT. For the latter, we used the
normalised Stokes 
$I$ and $V$ spectra that are stored as telescope data products in the
CFHT Archive. In the case of HARPSpol, the normalisation of the reduced
spectra retrieved from the ESO Archive was carried out by Ilya Ilyin
by application of the procedure described by
\citet{2013AN....334.1093H}.

To ensure that the mean longitudinal magnetic field values that are
determined from these spectra are as consistent as possible with the
published data of \citet{1997A&AS..124..475M} and of
\citet{2017A&A...601A..14M}, we applied the following procedure. We
used the same programme as these authors to measure a selected set of
20 Fe~{\sc i}  diagnostic lines with wavelengths comprised between
5400 and 6800\,\AA. This is approximately the range covered by
the CASPEC spectra of \citet{2017A&A...601A..14M}, whose analysis was
also based on a set of Fe~{\sc i} lines. Here, we used this same set
of lines, augmented by a number of Fe~{\sc i} lines that
were identified as suitable for the considered measurements as part of
the study 
of \citet{2006A&A...453..699M}. The addition of these diagnostic lines
was made possible in part by the fact that the spectral resolutions of
the HARPSpol spectrum ($R\approx115,000$) and of
the ESPaDOnS spectra ($R\approx65,000$) are higher than those of
the CASPEC 
spectra ($R\approx18,000$ to 39,000), so
that line blending is less of an issue in the former than in the latter.

All 21 values of the mean longitudinal field of HD~166473 resulting
from the observations and analyses described above are listed
in Table~\ref{tab:Hz}. The columns give, in order, the Heliocentric
Julian Date of mid-observation, the value $\Hz$ of the mean
longitudinal magnetic field and its uncertainty $\sigma_z$, and the
source of the measurement. For the FORS-1 measurements of
\citet{2015A&A...583A.115B}, we adopted the $\Hz$ values based on the
analysis of the metal lines.

\section{Variability and rotation period}
\label{sec:per}
\begin{figure}
\resizebox{\hsize}{!}{\includegraphics[angle=270]{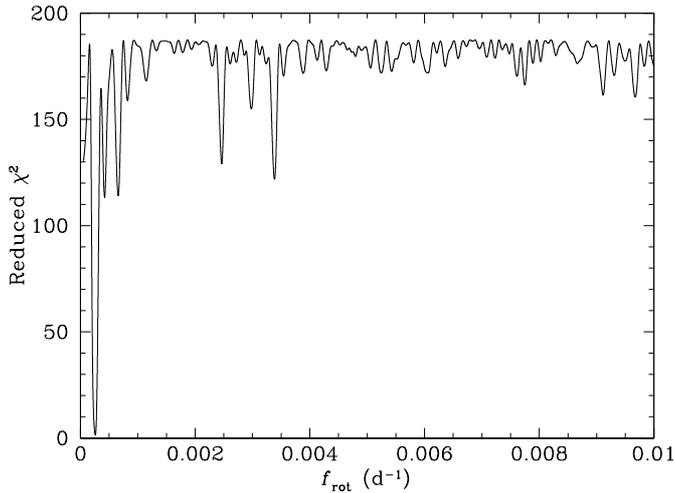}}
\caption{Periodogram of the variations of the mean longitudinal magnetic field
  of HD~166473. The ordinate is the reduced $\chi^2$ of a fit of
  the $\Hm$ measurements by a cosine wave, with the frequency given in
  abscissa. }
\label{fig:periodo}
\end{figure}

To determine the rotation  period of HD~166473, we fitted the
measurements of its mean magnetic field modulus by a cosine
wave, progressively varying the period of this wave, in search
of the value of the period that minimises the reduced $\chi^2$ of the
fit. As shown in Fig.~\ref{fig:periodo}, the application of
  this procedure unambiguously indicates that the best value of the
  period is of the order of 3836\,d.  

The phase variation curve of $\Hm$ for this value of the period is
shown in Fig.~\ref{fig:bmcurve}. The time elapsed between the
  first determination of the mean magnetic field modulus of HD~166473
  by \citet{1997A&AS..123..353M} 
and our most recent observation of the star is 9854\,d, or $\sim$2.6
rotation periods. The descending branch of the variation curve, between phases
0.05 and 0.25, is particularly interesting, as it features
measurements obtained during three consecutive cycles. The data
represented by circles (filled or open) are the oldest ones; the
filled triangle corresponds to a measurement obtained one cycle later,
and the filled squares identify observations acquired two cycles
later. By increasing the trial value of the period to 3866\,d, the
squares are systematically shifted below the best fit curve, but
within $\sim$1\,$\sigma$ of it. Conversely, for a trial period value
of 3806\,d, the squares all appear above the best fit curve, again
within $\sim$1\,$\sigma$ of it. Outside this period range, the
systematic differences between $\Hm$ measurements from different
cycles becomes too large to be accounted for only by measurement
uncertainties. This constrains the uncertainty affecting the
  derived value of the period:
\begin{equation}
\Prot=(3836\pm30)\,\mathrm{d}.
  \label{eq:Prot}
\end{equation}
The shape of the variation curve of HD~166473 appears
  remarkably close to 
  a pure cosine wave (see Sect.~\ref{sec:magfield}), but the accuracy
  of the derived value of the period only depends on the
    reproducibility of the variation curve from cycle to cycle,
    regardless of its exact shape. 
  
On the other hand, the determination of the period rests on the
implicit assumption that the $\Hm$ values that are derived from the
analysis of spectra recorded with different instruments, or instrument
configurations, are mutually consistent. Consideration of
Fig.~\ref{fig:bmcurve} gives strong support to the validity of this
assumption. The latter is also borne out, more generically, by the
overall consistency of all the $\Hm$ measurements analysed in our
recent studies of other Ap stars that rotate extremely slowly
\citep{2016A&A...586A..85M,2019A&A...624A..32M,2019A&A...629A..39M}. With
eight different 
instrumental configurations used to carry out the mean magnetic field
modulus measurements listed in Table~\ref{tab:Hm}, the consistency
between all of them illustrates the robustness of the $\Hm$
determinations obtained from the analysis of the \Feline\ line against
systematic errors of instrumental origin. Exceptions certainly occur
in some cases \citep[see e.g.][]{1997A&AS..123..353M}, but they are
rare and their impact on the conclusions that can be drawn about the
stellar physical properties from affected studies tends to be moderate. 

It should be noted that the rotation period of HD~166473 is longer
than the time base spanned by the magnetic measurements published by
\cite{2007MNRAS.380..181M}. The additional $\Hm$ determinations
  obtained by \citet{2014IAUS..302..274S} may at most have extended this
  time base by a limited fraction of the rotation period, leaving its value
  poorly constrained in the study of these authors.
Thus, our analysis provides the
first accurate determination of this value.

\section{Magnetic geometry and surface inhomogeneities}
\label{sec:magfield}
\subsection{Magnetic variation curves}
\label{sec:magcurves}
\begin{figure}
\resizebox{\hsize}{!}{\includegraphics{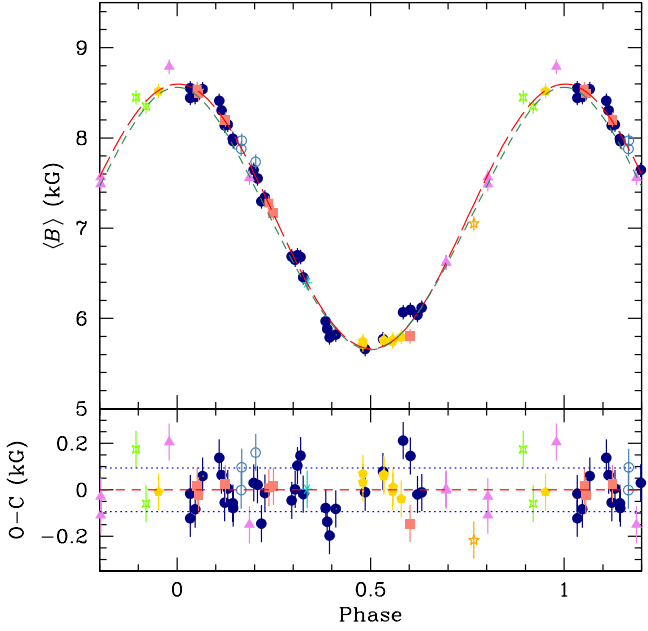}}
\caption{\textit{Upper panel:\/} Mean
  magnetic field modulus of HD~166473 against rotation
  phase. The different symbols identify the instrumental
  configuration from which the $\Hm$ value was obtained, as follows:
  filled circles (dark blue): CAT + CES LC; asterisk (turquoise): CAT
  + CES LC, lower resolution; open circle (steel blue):
  CAT + CES SC; filled
  triangles (violet): CFHT + Gecko \citep[all previous symbols
  identical to][]{2017A&A...601A..14M}; five-pointed open star
  (orange): 3.6\,m + CES VLC; four-pointed open stars (light
  green): 
  UT2 + UVES; filled pentagons (yellow): 3.6\,m + HARPS; filled
  squares (salmon): CFHT + ESPaDOnS. The
long-dashed line (red) is the best fit of the observations by a cosine wave
 -- see Eq.~(\ref{eq:bmfit}). The short-dashed
line (dark green) corresponds to the superposition of low-order multipoles
discussed in Sect.~\ref{sec:magfield}.
\textit{Lower panel:\/}
Differences $\mathrm{O}-\mathrm{C}$ between the individual $\Hm$
measurements and the best fit curve, against rotation phase. The
dotted lines (blue) correspond to $\pm1$~rms deviation of the
observational data about the fit (red dashed line). The symbols are
the same as in the upper panel.}
\label{fig:bmcurve}
\end{figure}

The observed variation of the mean magnetic field modulus of HD~166473
can be well represented by a cosine wave. The best least-squares fit
solution for $\Prot=3836$\,d is:
\begin{eqnarray}
\Hm(\phi) [G]&=&(7130\pm138)\nonumber\\
&+&(1467\pm19)\,\cos\{2\pi\,[\phi-(0.002\pm0.002)]\}\nonumber\\
         &&\hspace{-3em}(\nu=53,\ \chi^2/\nu=1.5),\label{eq:bmfit}
\end{eqnarray}
where the field strength is expressed in Gauss, $\phi=({\rm
  HJD}-{\rm HJD}_0)/\Prot$ (mod 1) and the adopted 
value of ${\rm HJD}_0=2448660.0$ corresponds to a maximum of
the mean magnetic field modulus, within the uncertainty of the phase;
$\nu$ is the number of degrees of 
  freedom, and $\chi^2/\nu$, the reduced $\chi^2$ of the fit. The
  fitted curve 
is shown in Fig.~\ref{fig:bmcurve}; the $\mathrm{O}-\mathrm{C}$
differences between the individual 
  measurements and this curve are also illustrated. The moderate value
  of the reduced $\chi^2$ is consistent with the views that the value
  adopted for the uncertainty affecting the derived $\Hm$ values is
  realistic, and that the actual shape of the $\Hm$ variation curve in
  HD~166473 is remarkably close to a pure cosine wave. This latter
  conclusion is particularly meaningful since the rotation cycle is
  sampled densely and almost uniformly by the available
  measurements. The scatter of the $\OC$ data points is
    rather uniform and does not show any obvious systematic differences
    between different instrumental configurations. This justifies the
    adoption of a single value of the $\Hm$ measurements errors,
    regardless of the spectral resolution or the S/N of the analysed
    spectra (see Sect.~\ref{sec:Hm}).

\begin{figure}
\resizebox{\hsize}{!}{\includegraphics{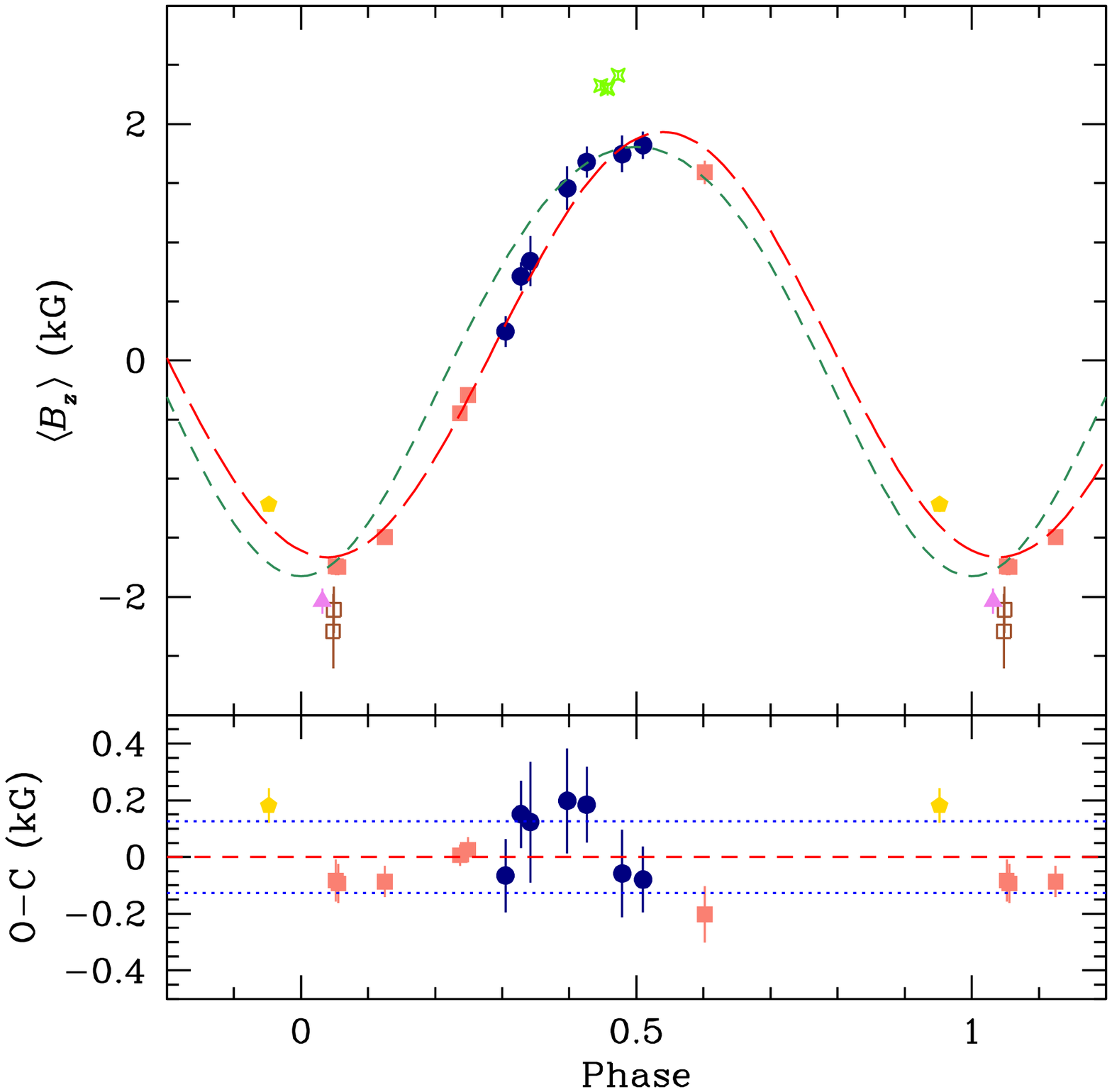}}
\caption{\textit{Upper panel:\/} Mean longitudinal magnetic field of
  HD~166473 against rotation 
  phase. The different symbols identify the sources of the $\Hz$
  values, as follows: 
  open squares (brown) or filled triangle
  (violet): \citet[CASPEC]{1997A&AS..124..475M}; filled circles (dark blue): 
  \citet[CASPEC]{2017A&A...601A..14M}; four pointed open stars (light green):
  \citet[FORS-1]{2015A&A...583A.115B}; filled pentagon (yellow): this
  paper (HARPSpol); filled squares (salmon): this
  paper (ESPaDOnS). The
long-dashed line (red) is the best fit of the CASPEC measurements of
\citet{2017A&A...601A..14M} and of the HARPSpol and ESPaDOnS
data by a cosine wave 
 -- see Eq.~(\ref{eq:bzfit}).
The short-dashed
line (green) corresponds to the superposition of low-order multipoles
discussed in Sect.~\ref{sec:magfield}.
\textit{Lower panel:\/}
Differences $\mathrm{O}-\mathrm{C}$ between the individual $\Hz$
measurements and the best fit curve, against rotation phase. The
dotted lines (blue) correspond to $\pm1$~rms deviation of the
observational data about the fit (red dashed line). The symbols are
the same as in the upper panel. The CASPEC measurements of
\citet{1997A&AS..124..475M} and the FORS-1 measurements are not shown 
in this panel as they were not included in the fit (see text).} 
\label{fig:bzcurve}
\end{figure}

The sampling of the variation curve of the mean longitudinal magnetic
field is sparser, with a gap of more than one third of a cycle between phases
0.60 and 0.95. Furthermore, as anticipated in Sect.~\ref{sec:Hz}, the
$\Hz$ values determined from FORS-1 observations that were obtained
around phase 0.45 show systematic differences with respect to the
$\Hz$ values based on CASPEC spectra recorded around the same
phase.
Furthermore, we consider the CASPEC-based measurements of the
  mean longitudinal magnetic field from \citet{1997A&AS..124..475M} to be
  less reliable than those from \citet{2017A&A...601A..14M}. Indeed,
  \citeauthor{1997A&AS..124..475M} used transitory configurations of
  the CASPEC spectrograph, which were not fully
  characterised. Moreover, two of their measurements (represented by
  open squares in Fig.~\ref{fig:bzcurve}), which have higher formal
  errors $\sigma_z$, were based on spectra taken with a configuration
  that was used only for a single, short observing run, and spanned a shorter 
  wavelength range than any other spectropolarimetric configuration of
  CASPEC.
Therefore, to compute the best-fit curve that
appears in Fig.~\ref{fig:bzcurve}, only the CASPEC data of
  \citet{2017A&A...601A..14M} were 
combined with the new $\Hz$ measurements that we obtained from the
analysis of HARPSpol and ESPaDOnS spectra.
Figure~\ref{fig:bzcurve}  also includes the 
representative points of the $\Hz$ measurements that
\cite{2015A&A...583A.115B} performed with FORS-1 and of
the CASPEC data of \citet{1997A&AS..124..475M}, but
these values were not included in the computation of the fit
parameters. 

 The best-fit solution that we derived is as follows:
\begin{eqnarray}
\Hz(\phi) [G]&=&(134\pm49)\nonumber\\
&+&(1798\pm51)\,\cos\{2\pi\,[\phi-(0.540\pm0.005)]\}\nonumber\\
         &&\hspace{-3em}(\nu=11,\ \chi^2/\nu=2.3).\label{eq:bzfit}
\end{eqnarray}
The fit is weighted by the inverse of the square of the uncertainties
of the individual measurements. 
The rather low value of the reduced $\chi^2$ suggests that the
measurement uncertainties are estimated correctly, that the CASPEC,
HARPSpol and 
ESPaDOnS $\Hz$ determinations are mutually consistent within these
uncertainties, and that the shape of the variation curve of the mean
longitudinal magnetic field modulus of HD~166473 does not differ
significantly from a cosine wave. Visual inspection of
Fig.~\ref{fig:bzcurve} confirms this conclusion, and there are no
indications of systematic differences between the $\Hz$ values
determined from the analysis of spectra taken with the three different
  instrumental configurations under consideration. Taking into account
  their higher formal errors, the 
  measurements of 
\cite{1997A&AS..124..475M} are also consistent with the data that were
included in the fit.

\subsection{Magnetic geometry}
\label{sec:maggeom}
An important
feature of the variation curve of the mean longitudinal magnetic field
of HD~166473 is that $\Hz$ undergoes sign
reversals. This implies that both poles of the star come alternately
into sight over a rotation cycle, a behaviour that is not predicted by
the preliminary model of the magnetic field structure of HD~166473
proposed by \citet{2003BSAO...56...25G}. Therefore, this model, which
was only constrained by mean magnetic field modulus measurements, must
be ruled out.

\begin{table}
\caption{Mean quadratic magnetic field measurements.}
\label{tab:Hq}
%\centering
\begin{tabular}{@{}@{\extracolsep{\fill}}crrl @{\extracolsep{0pt}}@{}}
\hline\hline\\[-4pt]
\multicolumn{1}{c}{JD}&\multicolumn{1}{c}{$\Hq$ (G)}&$\sigma_{\rm q}$ (G)&Reference\\[4pt]
\hline\\[-4pt]
 2448782.707& 9935&2091&\protect{\citet{1997A&AS..124..475M}}\\
 2448845.631&10774& 772&\protect{\citet{1997A&AS..124..475M}}\\
 2448846.696&11210& 573&\protect{\citet{1997A&AS..124..475M}}\\
 2449830.876& 7434& 282&\protect{\citet{2017A&A...601A..14M}}\\
 2449916.826& 7556& 155&\protect{\citet{2017A&A...601A..14M}}\\
 2449972.637& 6482& 316&\protect{\citet{2017A&A...601A..14M}}\\
 2450183.873& 6840& 211&\protect{\citet{2017A&A...601A..14M}}\\
 2450294.775& 6630& 348&\protect{\citet{2017A&A...601A..14M}}\\
 2450497.889& 6188& 160&\protect{\citet{2017A&A...601A..14M}}\\
 2450616.882& 6183& 162&\protect{\citet{2017A&A...601A..14M}}\\
 2454336.612& 6849& 128&This paper (HARPS)\\
 2454338.630& 6886& 110&This paper (HARPS)\\
 2454544.782& 6796& 112&This paper (HARPS)\\
 2454545.866& 6733& 126&This paper (HARPS)\\
 2454633.768& 6770& 120&This paper (HARPS)\\
 2454634.852& 6951& 147&This paper (HARPS)\\
 2454716.624& 6872& 122&This paper (HARPS)\\
 2456148.655& 9590& 395&This paper (HARPSpol)\\
 2456531.773&10200& 130&This paper (ESPaDOnS)\\
 2456547.732&10110&  98&This paper (ESPaDOnS)\\
 2456813.014& 9709& 122&This paper (ESPaDOnS)\\
 2457239.836& 9057& 122&This paper (ESPaDOnS)\\
 2457287.712& 8964& 161&This paper (ESPaDOnS)\\
 2458642.993& 8394& 172&This paper (ESPaDOnS)\\[4pt]
\hline
\end{tabular}
\end{table}

It is noteworthy
that, to the achieved precision, which is very high for the mean
magnetic field modulus, both the $\Hm$ and $\Hz$ curves are
mirror-symmetric about rotation phases $\sim$0 and
  $\sim$0.5. This is consistent 
with a magnetic field distribution that is symmetric about an axis
passing through the centre of the star. However, the least-squares
fits of the $\Hm$ and $\Hz$ variation curves indicate that the phases
of the extrema of these two field moments are shifted with
respect to each other by a small but formally significant
  amount.
Further confirmation of the significance of
this shift needs to be sought in the future by completing the
phase coverage of the $\Hz$ measurements. In any event, it is
  considerably smaller than the shift between the $\Hz$ and $\Hm$
  extrema that is observed in other super-slowly rotating Ap stars,
  such as HD~18078 \citep{2016A&A...586A..85M} or HD~50169
  \citep{2019A&A...624A..32M}. 

Comparison of the variation curves of the mean magnetic
field modulus and of the mean longitudinal magnetic field definitely
indicates that 
the field intensity is stronger over the part of the surface of the
star that is seen at the phase of the negative extremum of $\Hz$, to
which we shall hereafter refer to as the negative magnetic pole,
than around the part of the stellar surface that is observed at the
phase of the positive $\Hz$ extremum (hereafter the positive
pole). This represents a clear indication that,
although it is nearly axisymmetric,
the geometrical structure of the magnetic field
of HD~166473 must depart considerably from a single dipole at the
centre of the star. This conclusion is also supported by the
large value of the ratio between the values of 
the extrema of $\Hm$, $q=1.52$. The latter rules out a simple
dipolar field 
geometry, for which the maximum possible value of $q$ is $\sim$1.25
\citep{1969ApJ...158.1081P}. 

To gain more insight, it is a good approximation to try to represent
the structure of the magnetic field of HD~166473 by a simple
axisymmetric model, such as the superposition of collinear dipole,
quadrupole and octupole \citep{2000A&A...359..213L}. Through the
same procedure as used by these authors, we found that the parameter
values of the model that best reproduces the observations are as
follows: $i=36^\circ\pm3^\circ$, $\beta=90^\circ\pm3^\circ$, $B_{\rm
  dipole}=(-8820\pm300)$\,G,  $B_{\rm quadrupole}=(-6580\pm300)$\,G,
and $B_{\rm octupole}=(3190\pm300)$\,G. 
The variation curves of $\Hm$ and $\Hz$ that are computed with
  these parameters are shown in Figs.~\ref{fig:bmcurve} and
  \ref{fig:bzcurve}. 
This simple field model does not include the toroidal field and
is only meant to obtain a preliminary 
estimate of the geometry of the star, that is, primarily, to constrain
the inclination of the rotation axis to the line of sight (angle $i$)
and the angle $\beta$ between the magnetic axis (defined as the common
axis of 
the collinear bipolar, quadrupolar and octupolar components of
  the classical poloidal field solution considered here) and the
rotation axis. As
usual, the angles $i$ and $\beta$ may be exchanged with no change in
the predicted curves. It can be noted that the values derived
  here for these angles are not significantly different from those
  obtained by \citet{2000A&A...359..213L}, even though their preliminary
  model of HD~166473 was based on an estimate of the period
  extrapolated from observations spanning only 2300\,d, and
  accordingly leaving a major gap in the phase coverage of the
  variations. Therefore, further comparison of this model with the one
  derived here is not meaningful.

\begin{figure}
\resizebox{\hsize}{!}{\includegraphics{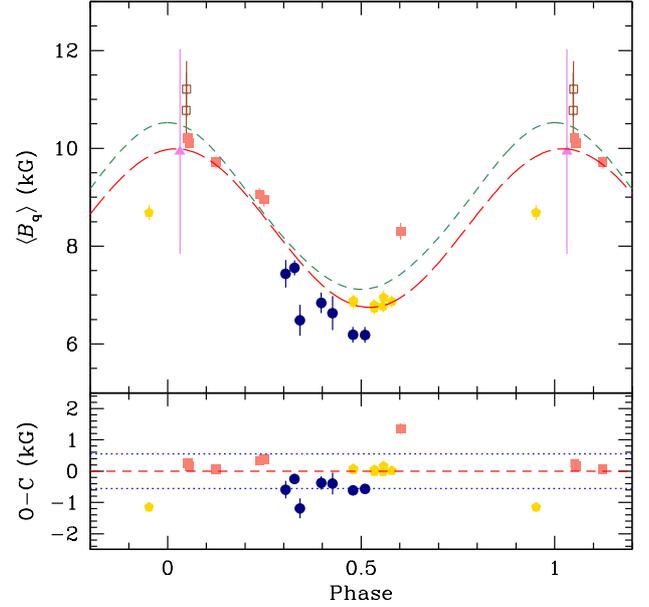}}
\caption{\textit{Upper panel:\/} Mean quadratic magnetic field of
  HD~166473 against rotation 
  phase. The symbols have the same meaning as in
  Fig.~\ref{fig:bzcurve}.
  The
long-dashed line (red) is the best fit of the CASPEC measurements of
\citet{2017A&A...601A..14M} and of the HARPS and ESPaDOnS
data of this paper by a cosine wave 
 -- see Eq.~(\ref{eq:bqfit}).
The short-dashed
line (green) corresponds to the superposition of low-order multipoles
discussed in Sect.~\ref{sec:magfield}.
\textit{Lower panel:\/}
Differences $\mathrm{O}-\mathrm{C}$ between the individual $\Hq$
measurements and the best fit curve, against rotation phase. The
dotted lines (blue) correspond to $\pm1$~rms deviation of the
observational data about the fit (red dashed line). The symbols are
the same as in the upper panel. The CASPEC measurements of
\citet{1997A&AS..124..475M} are not shown 
in this panel as they were not included in the fit (see text).} 
\label{fig:bqcurve}
\end{figure}

The model parameters are valuable on a statistical
  basis. However, the simple representation that it provides is not
  meant to be physically realistic, so that the uncertainties given
  for the parameters are only formal. Neither should one expect this
  geometrical model to yield a physically meaningful magnetic map of
  the stellar 
  surface. The observations that we have
  at our disposal until now are too incomplete and too inhomogeneous to
  justify an attempt at building a more realistic physical model. We plan to do
  so in the future, once we have acquired a complete set of data of
  high and uniform quality, which is in progress but whose completion
  still requires a few more years, 
  given the length of the rotation period of HD~166473.

\subsection{Mean quadratic magnetic field}
  \label{sec:Hq}
\citet{2007MNRAS.380..181M} and \citet{2017A&A...601A..14M} also
presented measurements of the mean quadratic magnetic field $\Hq$ of
HD~166473, which were used by \citet{2000A&A...359..213L} to constrain
their tentative model, based on an assumed value of the period, of the
geometric structure of the magnetic field of this star. The mean
quadratic magnetic field is the square root of the sum of two field
moments: (1) the average over the visible stellar disk of the square
of the modulus of the magnetic vector, $\Hsq$, and (2) the average
over the visible stellar disk of the square of the component of the
magnetic vector along the line of sight, $\Hzsq$. Both averages are
weighted by the local emergent line intensity.

As part of the analysis of the HARPSpol and ESPaDOnS spectra that was
carried out to determine the mean longitudinal magnetic field, we
also derived values of the mean quadratic magnetic field. Additional
determinations of this field moment were carried out from the HARPS
spectra recorded in natural light from which $\Hm$ measurements were
obtained (see Sect.~\ref{sec:Hm}).  The resulting fourteen 
newly derived values of $\Hq$ are presented in
Table~\ref{tab:Hq}, together with the ten 
measurements that had been previously published. The columns give, in
order, the Heliocentric 
Julian Date of mid-observation, the value $\Hq$ of the mean
quadratic magnetic field and its uncertainty $\sigma_{\rm q}$, and the
source of the measurement. The corresponding
phase diagram is shown in Fig.~\ref{fig:bqcurve}.

The determination of $\Hq$ from the HARPS and ESPaDOnS spectra was
carried out as described in detail by \citet{2017A&A...601A..14M}. In
particular, we took advantage of the availability of six ESPaDOnS
spectra obtained with the same configuration to derive the
non-magnetic contributions to the second-order moments of the Stokes
$I$ line profiles from consideration of the average of the profiles
at the six epochs of observation. The same approach was
applied to the eight HARPS observations. 

The determination of the mean quadratic magnetic field involves a
multiple linear regression analysis to untangle the contributions of
various broadening effects to the overall Stokes $I$ line widths, as
measured by the second-order moments of their profiles about their
centres. As noted by \citet{2006A&A...453..699M}, the achievable
precision on the derived value of $\Hq$ may be limited by the
occurrence of some crosstalk between the Doppler and Zeeman terms of
the regression equation. The Doppler term includes the contribution of
the instrumental profile, which depends on the spectral
resolution. This may lead to systematic differences between the values
of $\Hq$ that are determined from spectra obtained with different
instruments. \citet{2017A&A...601A..14M} reports that the values $\Hq$
that he derives from CASPEC spectra ($R\approx39,000$) are not fully
consistent with the values obtained from EMMI spectra
($R\approx70,000$) by \citet{2006A&A...453..699M}, for some stars in
common. The consideration of Fig.~\ref{fig:bqcurve} suggests that
there may also be systematic differences between the quadratic field
determinations carried out with each of the three instruments:
  CASPEC, HARPS and ESPaDOnS. In particular, the ESPaDOnS data point
  at phase 0.6 
seems inconsistent with the CASPEC and HARPS measurements
in the 
phase range 0.3--0.6. The HARPS determination at phase 0.95 also seems
difficult to reconcile with the CASPEC and ESPaDOnS field values from
the phase interval 0.0--0.1. These systematic effects
appear all the more likely since the observations with each of the
three instruments 
sample different parts of the stellar rotation cycle, which are
characterised by very 
different values of the mean magnetic field modulus. That the
crosstalk between the Doppler and Zeeman terms of the regression
equation is different in these these three phase intervals seems very
plausible. Since there are no other stars until now for which sets of
$\Hq$ measurements have been obtained with CASPEC, HARPS and ESPaDOnS, 
we cannot draw further conclusions about the possible different systematic
effects that may affect the determinations of the mean quadratic
magnetic field with these three instruments.

Accordingly, we do not believe that the different sets of mean
quadratic magnetic field measurements available for HD~166473 can be
reliably combined to constrain the geometrical structure of the
magnetic field of this star. Therefore, we computed a fit of the
variation of $\Hq$ by a cosine wave only for illustrative purposes:
\begin{eqnarray}
\Hq(\phi)&=&(8368\pm125)\nonumber\\
&+&(1621\pm146)\,\cos\{2\pi\,[\phi-(0.022\pm0.023)]\}\nonumber\\
         &&\hspace{-3em}(\nu=18,\ \chi^2/\nu=10.7).\label{eq:bqfit}
\end{eqnarray}
This fit, which is weighted by the inverse of the square of the
uncertainties of the individual measurements, is based on the CASPEC
data from \citet{2017A&A...601A..14M} and on the HARPS and ESPaDOnS
$\Hq$ determinations of this paper. The value of the reduced $\chi^2$
is somewhat high, which lends support to the suspicion that the mean
quadratic magnetic field values derived from spectra taken with the different
instruments may not be fully consistent. The consideration of the
distribution of the data points about the best-fit curve in
Fig.~\ref{fig:bqcurve} further strengthens this interpretation. It
should be noted that the addition of the first harmonic to the fitted
function does not improve the quality of the
fit. Figure~\ref{fig:bqcurve} also shows the $\Hq$ variation curve
predicted by the simple multipolar model discussed above.The model
provides a rough but reasonable approximation of the
observed variations of the quadratic field, whether or not the
measurements of this field moment are used to constrain it. In
particular, that the model curve tends to be shifted 
towards higher values with respect to the measurements is not
inconsistent with the observation that, frequently, the derived values
of $\Hq$ tend to slightly underestimate the actual magnetic field
\citep{2006A&A...453..699M,2017A&A...601A..14M}.

\begin{figure}
\resizebox{\hsize}{!}{\includegraphics{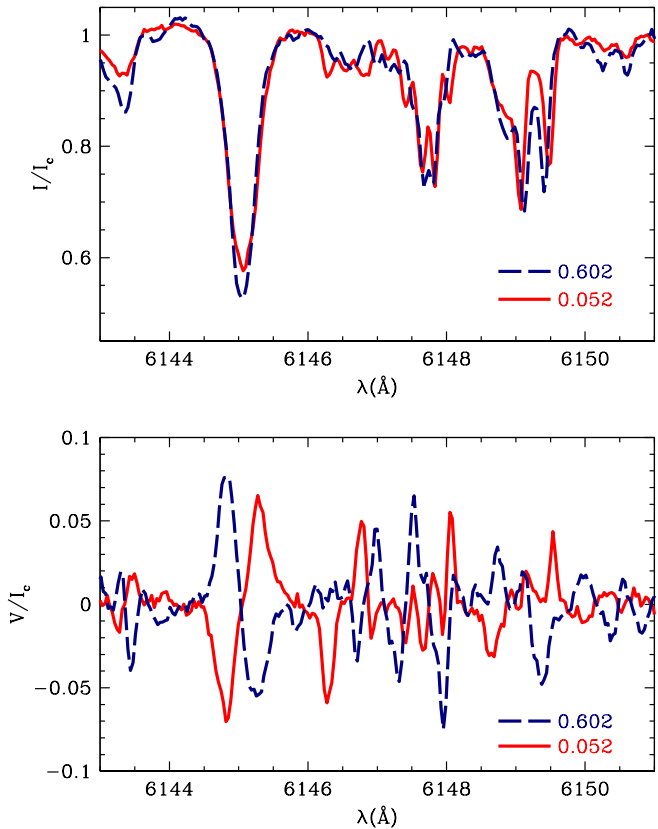}}
\caption{Part of the region of the spectrum of HD~166473 that is common to all
  the high-resolution spectra from which the mean magnetic field modulus
  measurements of Table~\ref{tab:Hm} were obtained. 
  Stokes $I$ (\textit{upper panel\/}) and Stokes $V$ (\textit{lower
    panel\/}) spectra, recorded at phases 0.052 (solid red line) and
  0.602 (dashed blue line) are shown; they are normalised to the local
  unpolarised continuum intensity $I_{\mathrm{c}}$. The main lines are
  \Ndline\ (not 
  fully resolved), \Felineb\  
  (with a quadruplet pattern), and \Feline\ (a doublet, blended on the
  blue side with an unidentified rare earth line).}
\label{fig:sp2}
\end{figure}

\begin{figure}
\resizebox{\hsize}{!}{\includegraphics{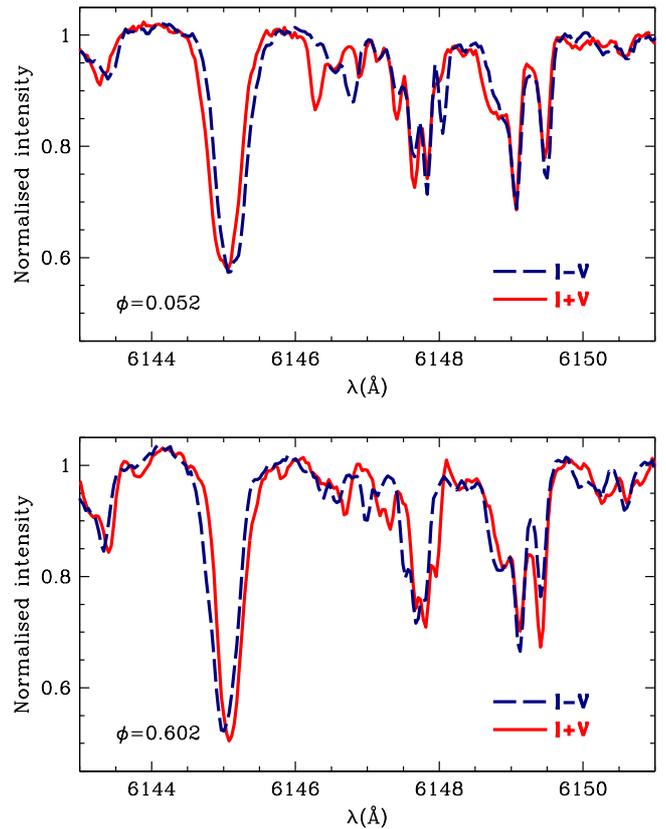}}
\caption{Part of the region of the spectrum of HD~166473 that is common to all
  the high-resolution spectra from which the mean magnetic field modulus
  measurements of Table~\ref{tab:Hm} were obtained. Spectra recorded
  at phases 0.052 (\textit{upper panel\/}) and 0.602 (\textit{lower
    panel\/}) are shown, in right circular polarisation ($I+V$; solid
  red line) and in left circular polarisation ($I-V$; dashed blue
  line). The main lines are \Ndline\ (not fully resolved), \Felineb\ 
  (with a quadruplet pattern), and \Feline\ (a doublet, blended on the
  blue side with an unidentified rare earth line).}
\label{fig:sp1}
\end{figure}

\subsection{Line intensity variations}
  \label{sec:linvar}
\begin{figure}
\resizebox{\hsize}{!}{\includegraphics[angle=270]{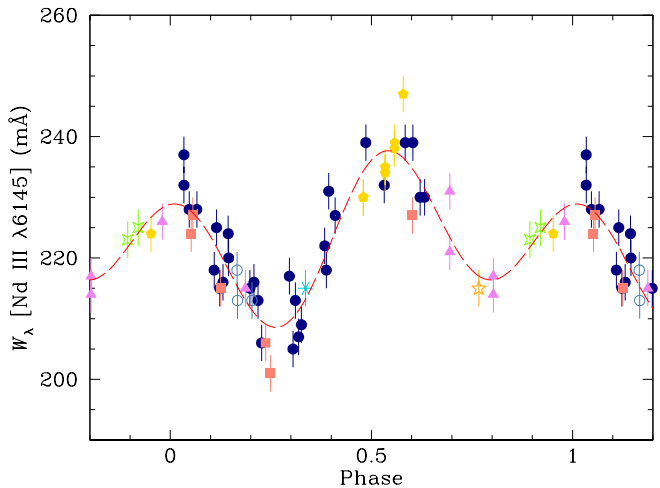}}
\caption{Equivalent width of the \Ndline\ line against rotation
  phase. The symbols have the same meaning as in
  Fig.~\ref{fig:bmcurve}. The
long-dashed line (red) is the best fit of the measurements by a
cosine wave and its first harmonic. It has been included to guide the
eye despite its lack of physical meaning. 
}
\label{fig:ndew}
\end{figure}

\begin{figure}
\resizebox{\hsize}{!}{\includegraphics[angle=270]{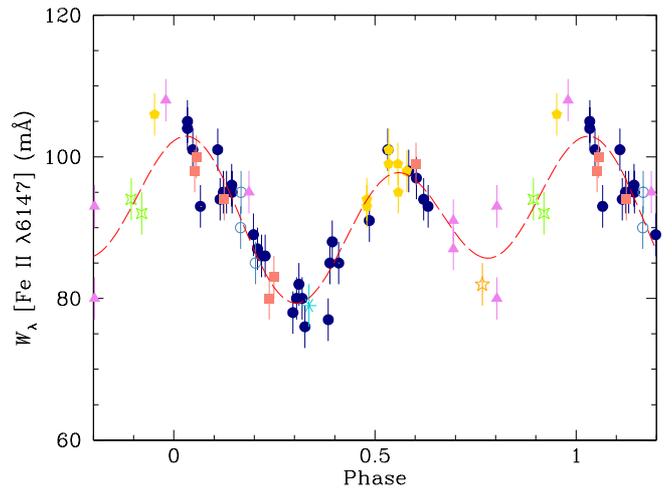}}
\caption{Equivalent width of the \Felineb\ line against rotation
  phase. The symbols have the same meaning as in
  Fig.~\ref{fig:bmcurve}. The
long-dashed line (red) is the best fit of the measurements by a
cosine wave and its first harmonic. It has been included to guide the
eye despite its lack of physical meaning. 
}
\label{fig:feew}
\end{figure}

With the diversity of instrumental configurations used for the
determination of the mean magnetic field modulus, only a narrow
wavelength range is common to all the recorded spectra. A segment
  of this common range is shown in Fig.~\ref{fig:sp2}. It includes
  two of the few 
spectral lines in this region that appear reasonably free from blends:
\Ndline\ and \Felineb. These two lines
show significant variations with rotation phase, as can be
seen in  Figs.~\ref{fig:ndew} and \ref{fig:feew}. Both lines
are strong: the overabundance of the Rare Earth Elements and the
moderate enhancement of the iron-peak elements in the photosphere of
HD~166473 were first reported by \citet{2000A&A...356..200G}. In
hotter Ap stars, the latter line arises almost entirely from a Fe~{\sc
  ii} transition of the same multiplet 74 as \Feline. However, in cool stars
such as HD~166473, it also includes a significant contribution from a
Fe~{\sc i} line at 6147.8\,\AA\ 
\citep{1992A&A...256..169M}. This does not represent an issue, as the
horizontal distributions of Fe\,{\sc i} and Fe\,{\sc ii} over the
surface of Ap stars are in general identical. While there may also be
a small contribution of the Ti~{\sc i}~$\lambda\,6147.76$ transition,
the main element responsible for the observed line at 6147.7\,\AA\ and
its variations is undoubtedly Fe. At the field strengths
present in HD~166473, magnetic desaturation of the spectral lines is
almost complete, so that the observed line intensity variations cannot
be assigned to differences in magnetic line intensification between
different parts of the stellar surface. Instead, they must reflect the
existence of horizontal abundance inhomogeneities across this surface. 

The Stokes $I$ and $V$ spectra shown in Fig.~\ref{fig:sp2} were
obtained 
  with ESPaDOnS close to the phases of the magnetic extrema. The
    Stokes $V$ plot
  clearly illustrates the reversal of the mean longitudinal
  magnetic field. In the Stokes $I$ tracings, one can easily see
    the greater
    wavelength separation of the split components of the resolved
    lines close to phase 0 than to phase 0.5.

The same observations can also be plotted in a different form, showing
the Stokes $I+V$ and $I-V$ spectra at the two phases of interest (see
Fig.~\ref{fig:sp1}), to emphasise the polarisation of the individual
split line components. This
is illuminating, in particular for the outermost
  $\sigma$ components of the \Felineb\ quadruplet at phase 0.602: the
  blue one is almost fully circularly polarised to the left, and the
  red one is almost fully circularly polarised to the right. (For a
  detailed description of the Zeeman pattern of the \Felinebb\
  transition, see \citealt{1990A&A...232..151M}.) The polarisations of
  these components are swapped, but less complete at phase 0.052. This
  is an indication of the presence of mixed field polarities on the stellar
  hemisphere that is visible around phase 0, while no field polarity
  change should occur over most of the
  hemisphere that is observed around phase 0.5. We expect to be able
  to confirm and fully characterise this through the detailed
  physical modelling that we plan to carry out once we have acquired
  spectra of the required quality with the necessary phase sampling
  (see Sect.~\ref{sec:maggeom}). The simple geometrical model
  presented here is unsuitable for such an endeavour.

The variation curves of the
equivalent widths of the lines \Ndline\ (Fig.~\ref{fig:ndew})
and \Felineb\ (Fig.~\ref{fig:feew}) both show two 
maxima and two minima per rotation cycle. The maxima occur close to the
phases of extrema of the magnetic field moments: that is, both lines
are stronger in the vicinity of the magnetic poles. The \Ndline\ line
is significantly stronger close to the positive pole than to the
negative one, while the \Felineb\ line seems rather stronger close to
the negative pole. The two equivalent widths minima, for each of the
two lines, also appear to differ from each other. This points at a
non-axisymmetric distribution of the elemental abundances over the
surface of HD~166473, despite the plausibly axisymmetric structure of
the magnetic field. This should not be regarded as a major
discrepancy, since contrary to what has been assumed for a long time, the
distribution of elemental abundances on the surfaces of the Ap stars
may show only loose correlations (or none at all) with the magnetic
structure \citep[e.g.,][]{2018A&A...609A..88R}.

On the other hand, Fe
lines were used for the
measurements of the mean magnetic field modulus and of the mean
longitudinal magnetic field from which we computed a model of the
structure of the field. Accordingly, the geometry that we derived is
actually a convolution of the actual structure of the magnetic field
with the distribution of the Fe line intensity across the
star. Untangling these two components would require elaborate
numerical modelling beyond the scope of the present study, for which
the observational data used in this analysis are
insufficient. Moreover, it is unclear to which extent and how uniquely
the untangling of the magnetic geometry and of the elemental abundance
distribution could be achieved since, in contrast with Ap stars with
rotation periods not exceeding a few weeks, there is no significant
rotational Doppler effect to shift the contributions of regions of
the stellar surface of different longitudes with respect to each other
in the observed, disk-integrated spectral line profiles.\\

\section{Discussion}
\label{sec:concl}
Following this study, HD~166473 becomes only the fourth Ap star with a
rotation period longer than 10 years for which magnetic field
measurements have been obtained over more than a full cycle.
With the addition of HD~166473, there are now eight Ap stars for which
rotation periods longer than 1000\,d have been accurately determined and
magnetic variations have been well sampled over a full rotation
period. These stars are listed, in order of decreasing period, in
Table~\ref{tab:stars}. The columns give, in order: the HD number of
the star, another identification, the spectral type according to
\citet{2009A&A...498..961R}, the rotation period, the reference from
which it is extracted, the rms mean longitudinal field $\Hzrms$
\citep[as defined by][]{1993A&A...269..355B}, the average over a
rotation cycle of the mean magnetic field modulus ($B_0$), the ratio
of $\Hzrms$ to $B_0$, the ratio $q$ of the extrema of the mean
magnetic field modulus, the ratio $r$ of the smaller (in absolute
value) to the larger (in absolute value) extremum of $\Hz$, and some
notes (SB1 and SB2 identify, respectively, single-lined and
double-lined spectroscopic binaries). Most information for each star was
extracted or computed from 
the sources from which the values of the periods were retrieved, and
from the references quoted in these sources. For HD~9996 and
HD~187474, additional magnetic field measurements from
\citet{2017A&A...601A..14M} were included in the computation of the
magnetic parameters.

 The $\Hz$ measurements from the quoted references were used to compute 
the values of $\Hzrms$ that appear in Table~\ref{tab:stars}.
Although possible in principle, determining the mean longitudinal
magnetic fields of stars in SB2 systems is considerably less
straightforward than in single stars. 
This has not been done yet in a systematic way for HD~59435.

For each star, the value that is given for the average of the mean magnetic field
modulus over a rotation cycle is that of the independent term $B_0$ of
the least-squares fit of the $\Hm$ measurements by a cosine wave, or
by a cosine wave and its first harmonic, as adopted in the quoted
reference. This value is undefined for HD~9996, since its spectrum
only shows resolved lines over a small fraction ($\sim$0.3) of its
rotation cycle \citep{2017A&A...601A..14M}. Outside this phase
interval, $\Hm$ cannot be determined. However, for most of the cycle,
its value must be considerably less than 3.5\,kG. Accordingly, the
ratio $q$ between the extrema of the mean magnetic field modulus of
HD~9996 may plausibly be close to 2.0, which is approximately the
largest value observed in any Ap star for which the $\Hm$ variation
curve is fully constrained. In this case, $B_0$ could be of the order
of 3.8\,kG, since the highest value of $\Hm$ that has been measured in
HD~9996 is $\sim$5.1\,kG.

\begin{table*}[!ht]
\caption{Ap stars with accurately determined rotation periods longer than
1000\,d.} 
\label{tab:stars}
\begin{tabular*}{\textwidth}[]{@{}@{\extracolsep{\fill}}rllrcrrcrrl}
%@{\extracolsep{0pt}}@{}}
\hline\hline\\[-4pt]
\multicolumn{1}{c}{HD}&Other id.
  &Sp. type&$\Prot$&Ref.&\multicolumn{1}{c}{$\Hzrms$}&\multicolumn{1}{c}{$B_0$}&\multicolumn{1}{c}{$\Hzrms/B_0$}&\multicolumn{1}{c}{$q$}&\multicolumn{1}{c}{$r$}&Notes\\
  &&&(d)&&(G)&(G)
\\[4pt] 
  \hline\\[-4pt]
50169&BD $-1$~1414&A3p SrCrEu&10600&1&1294&5076&0.25&1.43&$-$0.96&SB1\\
9996&HR 465&B9p CrEuSi&7937&2&713&--&--&--&$-$0.35&SB1\\
  965&BD $-0$~21&A8p SrEuCr&6030&3&775&4253&0.18&1.00&$-$0.50&\\
166473&CoD $-37$~12303&A5p SrEuCr&3836&4&1682&7131&0.24&1.52&$-$0.99&roAp\\
94660&HR 4263&A0p EuCrSi&2800&5&1911&6232&0.31&1.06&0.92&SB1\\
187474&HR 7552&A0p EuCrSi&2345&6&1470&5417&0.27&1.27&$-$0.98&SB1\\
59435&BD $-8$~1937&A4p SrCrEu&1360&7&--&3036&--&1.87&--&SB2\\
18078&BD $+55$~726&A0p SrCr&1358&8&692&3450&0.20&1.81&$-$0.92&SB1\\[4pt]
  \hline
\end{tabular*}
\tablebib{
  (1)~\citet{2019A&A...624A..32M};
  (2)~\citet{2012AcA....62..297B};
  (3)~\citet{2019A&A...629A..39M};
  (4)~This paper;
  (5)~\citet{2017A&A...601A..14M};
  (6)~\citet{1991A&AS...89..121M};
  (7)~\citet{1999A&A...347..164W};
  (8)~\citet{2016A&A...586A..85M}.}
\end{table*}

 Among the stars of Table~\ref{tab:stars}, HD~166473 has the strongest
mean magnetic field modulus, as characterised by its average value
over a rotation cycle, $B_0=7131$\,G. This value is close to, but
below, the 7.5\,kG value that seems to represent an upper limit to the
field strengths for Ap stars that rotate extremely slowly
($\Prot\ga150$\,d;
\citealt{1997A&AS..123..353M,2017A&A...601A..14M}). As mentioned in
Sect.~\ref{sec:intro}, among the roAp stars, HD~154708 definitely has
a stronger magnetic field than HD~166473. The case of HD~92499 is less
clearcut, as only a few measurements of its mean magnetic field
modulus have been obtained, and its variation is not well
characterised. The $\Hm$ values that have been determined for this
star at various epochs
\citep{2007MNRAS.378L..16H,2008MNRAS.389..441F,2010MNRAS.404L.104E},
which range from 8.2 to 8.5\,kG,
are greater than the average value of the mean magnetic field modulus
of HD~166473 over its rotation cycle, $B_0=7.1$\,kG, but they do not exceed
its value at the phase of maximum, 8.6\,kG. The average value of $\Hm$
over the rotation cycle of HD~92499, and how it compares with
HD~166473, is still unknown. Notwithstanding, HD~166473 is definitely
one of the most strongly magnetic roAp stars. 
Moreover, it is at present the only roAp star with a rotation
period longer than 1000\,d for which the accurate value of this period
has been determined and the magnetic variations have been fully
characterised. However, several other roAp stars definitely have
rotation periods of several years, which the observations that have been
obtained until now do not cover entirely yet. 

Within the framework of the extensive, parameter-based,
  statistical study of \citet{2017A&A...601A..14M},
HD~166473 does not stand out as
particularly remarkable in any respect. The ratios $q$ and $r$ of the
extrema of its mean magnetic field modulus and of its mean magnetic
longitudinal field, and the relation between the averages of these
field moments over a rotation cycle as characterised by the ratio
$\Hzrms/B_0$, are within the typical ranges
\citep{2017A&A...601A..14M}. That the variation curves of $\Hm$
and of $\Hz$ show no significant departure from harmonicity
differentiates HD~166473 from the other stars of Table~\ref{tab:stars},
except maybe HD~965 (for which no significant variation of $\Hm$ is
detected). However, such a behaviour is rather common in shorter
period stars \citep{2017A&A...601A..14M}. Moreover, the phase
difference (0.538) between the cosine fits to the $\Hz$ and $\Hm$
measurements is well within one of the ``normality'' bands of Fig.~11
of \citet{2017A&A...601A..14M}. 

More generally, consideration of Table~\ref{tab:stars} as a whole is
intriguing in several respects. First of all, the spectral types of
the stars in this table range from B9p to A8p, encompassing most of
the temperature interval in which classical Ap stars are found, except
for its lower end. This is rather surprising, at it is generally
accepted that lower mass and lower temperature Ap stars rotate in
average slower than their more massive, hotter counterparts
\citep[see Fig.~6 of][]{2017MNRAS.468.2745N}. One would rather expect
the majority of the most slowly rotating Ap stars to have late A or
early F spectral types. That this is not the case for those extremely long
period Ap stars of which a full rotation cycle has been covered by the
observations obtained until now may be coincidental, due to the small
number of these stars. The same explanation probably accounts
for the high fraction of spectroscopic binaries in
Table~\ref{tab:stars}. Indeed, out of the eight stars listed in this
table, six are spectroscopic binaries. This represents a fraction
considerably greater than the overall rate of occurrence of binarity
among Ap stars, which is of the order of 50\%\ according to the most
recent estimates \citep[][and references
therein]{2017A&A...601A..14M}. All the binaries listed in
Table~\ref{tab:stars} have orbital periods of several hundred days
(for their exact values, see \citealt{2017A&A...601A..14M}), so
that tidal interaction between the components must be negligible.

However, the most significant feature of Table~\ref{tab:stars} may be
the distribution of the values of the ratio $r$ between the extrema of
the mean longitudinal magnetic field. It is remarkable that $r<-0.90$
for half the stars of this table, and that $r<0.00$ for all of them
but HD~94660. Although measurements of $\Hz$ covering a full rotation
cycle have not been obtained for HD~59435, and no individual values
have been published, the range covered by the existing measurements,
as reported by \citet{2008AstBu..63..139R}, definitely indicates that
$r$ is negative. The distribution of the values of $r$ in
Table~\ref{tab:stars} can be compared with their distribution for the
Ap stars with resolved magnetically split lines for which $\Hz$
measurements well distributed over a rotation cycle are available, as
listed in Tables~13 and 14 of \citet{2017A&A...601A..14M}. Excluding
the stars also listed in Table~\ref{tab:stars} of the present
paper, these two tables contain 6 stars for which $r<0$ and 18 stars
for which $r>0$. The rotation periods of all of them are shorter than
1000\,d. For only one of them, HD~200311, $r<-0.90$. Even though
caution is still called for about the statistical significance of the
conclusions that can be drawn from a sample of only eight well studied
Ap stars with $\Prot>1000$\,d, the contrast in the rates of occurrence
of negative values of $r$, especially large ones,
between this group and the shorter period stars from the sample of
\citet{2017A&A...601A..14M}, is striking. Large (in absolute value)
negative values of $r$ are indicative of values close to $90^\circ$ of
one of the angles $i$ or $\beta$. There is no reason to expect any
difference between the stars with periods longer or shorter than
1000\,d in the distribution of the
inclination angles $i$ of the rotation axes to the line of sight. Thus,
any difference in the distribution of the $r$ values between the two
groups must arise from different distributions of the inclination of
the magnetic axes with respect to the rotation axes. Specifically, the
high rate of occurrence of mostly large negative values of $r$ in
Table~\ref{tab:stars} strongly suggests that the angle between the
magnetic and rotation axes systematically tends to be large in the
most slowly rotating Ap stars. The existence of such a trend was
already proposed by \citet{2017A&A...601A..14M}. The evidence
presented here strengthens the case.

With the growing number of extremely slowly rotating Ap stars whose
periods have been accurately determined and whose magnetic field variations
have been fully characterised over a full cycle, there is little doubt
left that, on an individual basis, these stars do not distinguish
themselves from faster rotating Ap stars in any other
respect. However, they may as a group have properties that
differentiate them from other subgroups of Ap stars. In this respect,
one of the most intriguing possibilities to which the available
evidence seems to point, albeit with limited statistical significance,
is a tendency for the angle between the magnetic and rotation axes to
be systematically large. Other aspects that deserve further
investigation include the rate of occurrence of extremely slowly
rotating Ap stars in binary systems, the confirmation and, possibly,
the further characterisation of the apparent lack of very strong
magnetic fields in the longest period Ap stars, and the possible
mass dependence of the occurrence of super slow
rotation. Knowledge of these various aspects and of the constraints that
they imply is essential for progress in the understanding of the
formation and evolution of the Ap stars and of their magnetic
fields. Therefore, it is critically important to acquire on a regular
basis new observations of the most slowly rotating Ap stars with a
view to building a set of such stars, whose rotation period is accurately
determined and whose magnetic field is fully characterised, that is
sufficiently populated to allow statistically significant conclusions
to be derived reliably. 

\begin{acknowledgements}
We thank the anonymous referee for calling our attention to the
  availability of the HARPSpol observation of HD~166473 in the ESO
  Archive, and for thought-provoking comments. We also thank Swetlana
  Hubrig and Ilya Ilyin for providing the normalised 
version of these reduced HARPSpol spectra that were used in this analysis.
VK and JDL acknowledge support from the Natural Sciences and Engineering
Research Council of Canada.
\end{acknowledgements}

\bibliographystyle{aa}
\bibliography{hd166473per_rev2}
\end{document}